\documentclass{article}

\usepackage{PRIMEarxiv}

\usepackage[utf8]{inputenc} 
\usepackage[T1]{fontenc}    
\usepackage{hyperref}       
\usepackage{url}            
\usepackage{booktabs}       
\usepackage{amsfonts}       
\usepackage{nicefrac}       
\usepackage{microtype}      
\usepackage{lipsum}
\usepackage{fancyhdr}       
\usepackage{graphicx}       
\graphicspath{{media/}}     

\title{FusionU-Net: U-Net with Enhanced Skip Connection for Pathology Image Segmentation
}

\author{
  Zongyi Li, Hongbing Lyu\\
  Zhejiang University \\
  Hangzhou, China\\
  \texttt{\{zongyi\_li, lhb\}@zju.edu.cn} \\
   \And
  Jun Wang \\
  Hangzhou City University \\
  Hangzhou, China\\
  \texttt{wjcy19870122@163.com} \\
}

\begin{document}
\maketitle

\begin{abstract}
    In recent years, U-Net and its variants have been widely used in pathology image segmentation tasks.
    One of the key designs of U-Net is the use of skip connections between the encoder and decoder, 
    which helps to recover detailed information after upsampling. While most variations of U-Net adopt the 
    original skip connection design, there is semantic gap between the encoder and decoder that can 
    negatively impact model performance. Therefore, it is important to reduce this semantic gap before 
    conducting skip connection.
    To address this issue, we propose a new segmentation network called FusionU-Net, which is based on U-Net 
    structure and incorporates a 
    fusion module to exchange information between different skip connections to reduce semantic gaps.
    Unlike the other fusion modules in existing networks, ours is based on a two-round fusion design 
    that fully considers the local relevance between adjacent encoder layer outputs and the need for 
    bi-directional information exchange across multiple layers. 
    We conducted extensive experiments on multiple pathology image datasets to evaluate our model 
    and found that FusionU-Net achieves 
    better performance compared to other competing methods. We argue our fusion module is more effective than 
    the designs of existing networks, and it could be easily embedded into other networks to further enhance the 
    model performance. Our code is available at: \url{https://github.com/Zongyi-Lee/FusionU-Net}
\end{abstract}


\section{Introduction}
Medical imaging segmentation is a crucial area of AI research with great
application value in areas such as computer aided diagnosis \cite{a:chromo} and image-guided surgery \cite{art:anwar2018medical} 
One of main chanllenges in this field is the lack of training data due to 
the fact that labeling medical images requires professional skills and also time-consuming.
Pathology images, in particular, with many small and densely distributed target areas, 
present additional difficulties to segmentation tasks.
As is pointed out by many researches, right inductive bias could help 
the model to generalize well on restricted training dataset \cite{art:inductive}\cite{art:geirhos2018imagenet}. 
While the inductive biases of Convolutional Neural Networks(CNNs) are 
locality and weights sharing, which are consistent with fact that for pathology images many features 
worth noticing are strongly locally related. In fact, CNNs do become dominant in such areas, 
and U-Net \cite{c:unet} is a specially successful one among numerous CNNs. 
U-Net is a typical encoder-decoder structure network, the shallow layers in the encoder part mainly work on capturing low-level 
features and the spatial semantic information grows richer as the layer progresses deeper. In the decoder part, 
upsampling is applied to restore the image to its full size, while skip connections are employed to combine 
coarse-grained features from deep layers with fine-grained features from shallow layers to aid in 
object details recovery. 
This design has led to U-Net's great success on numerous segmentation tasks and has also inspired many researches. 
However, most of these studies have only focused on improving the encoders or decoders but not altering 
the original skip connection design.

Recently, researchers have noticed the semantic gap between the encoder and decoder \cite{a:multi-resunet},
and \cite{c:uctransnet} carefully examined the contribution of different skip connections in U-Net and found that 
simple skip connections may not always be helpful and sometimes could even harm model performance. 
Based on this finding, they proposed a CCT module to fuse different feature maps of skip connections.
Their fusion method mainly focuses on channel-wise information but disregards the spatial relevance 
between adjacent layer outputs.
Tracing back to the forward process of U-Net, it is evident 
that convolutions are the dominant computing operations, while the characteristic of convolutions 
guarantees the feature maps before and after convolutions are strongly locally relevant. 
With above analysis, it is natural to deduce that one of the keys to bride the semantic gap among different layer outputs is to accurately 
capture the local relevance. To this end, we propose a new network called FusionU-Net, 
which uses U-Net as base structure and adopts a fusion module to apply feature fusion on skip connections. 

Unlike the design of previous works, we carefully considered the \emph{relevance of adjacent skip connections} and 
implemented a different approach for the feature fusion. Our fusion design is based on the following 
considerations: 1) The 
feature maps produced by the deep and shallow layers of the encoder are  inherently different while adjacent layer outputs 
have stronger relevance. Therefore, we only perform feature fusion between adjacent layers. 2) Our model should enable the information to be 
exchanged between any two encoder layer outputs. To achieve this we propose two-round fusion design. In the first round, we operate feature fusion 
between each pair of adjacent feature maps from shallow to deep encoder layers. In the second round, we operate in reverse order.
This allows information to still be exchanged between feature maps with multiple layers in between, even though they are not explicitly fused.
Additionaly the two-round design also enables the information to be switched bi-directionally. 3) The feature map of adjacent encoder layers 
are locally relevant, and the locality is also important characteristic of pathology images, thus we design a new method to 
fuse adjacent two feature maps together with fully considerations about keeping the feature map local adjacency, the details of which will be discussed 
in later section.  

Our model has a clear and simple structure and requires significantly fewer training parameters compared with other 
state-of-the-art methods. To test our model, we conduct extensive experiments on three datasets  and results strongly supports the 
superiority of our model. We argue that our fusion design represents a more effective and reasonable way and our fusion module 
could be easily embedded into other networks to further boost the performance of segmentation.

\section{Related Works}
\label{sec:related works}
\subsection{U-Net based Networks}
As various studies and application results having proved the effectiveness of U-Net design, many  networks base on U-Net have been proposed. 
H-DenseUNet \cite{a:denseunet} is a typical U-Net variation that learns from DenseNet \cite{c:densenet}, it incorporates a hybrid feature 
fusion(HFF) layer to fuse 2D and 3D features and achieves 
impressing results on liver segmentation task. \cite{a:r2unet} designs R2U-Net, it employs U-Net as the base structure and 
applies recurrent residual block in decoder part for performance improvement. AttentionU-Net \cite{a:attnunet} 
aims at suppressing irreverent regions by utilizing a attention gate module, the attention gate could effectively highlight specific local features 
while introducing acceptable extra costs. ResU-Net \cite{a:resunet} uses residual units to 
make it easier to train deeper networks. Based on ResU-Net, \cite{c:resunet++} adopted some new technologies such as squeeze and excitation 
blocks \cite{c:senet}, Atrous Spatial Pyramidal Pooling(ASPP) \cite{a:deeplabv3} and came up with a new model called ResU-Net++. 
Recently, with the great success of Vision Transformer(ViT) \cite{a:vit} and Swin-Transformer \cite{c:swintransformer}, 
people have started to recognize the potential of Transformer on computer vision tasks. TransUNet proposed by \cite{a:transunet} 
is the first model applying Transformer on U-Net structure for medical segmentation task, and they also use CNNs to extract low-resolution 
features since they believe the Transformer lay too much attention on global texture and tends to lack  detailed localization 
information. Later proposed Swin-UNet \cite{c:swinunet} is a pure Transformer-based U-Net like network and it keeps the skip connections 
between the encoder and decoder.

\subsection{Modification on skip connections}
The aforementioned works primarily focus on the modifying encoder or decoder modules of U-Net in order to enhance the ability 
of feature extraction. However, they ignore the fact that as a 
key part of U-Net design, better processed skip connections could also greatly assist the model performance.
 \cite{a:multi-resunet} noticed the semantic gap between the encoder and decoder, and they 
chose to use ResPath to enhance skip connection to alleviate the gap. But their ResPath is operated for each skip connection individually, 
not considering exchanging information between layers.  
Despite the design deficiencies, their work pointed out that the original skip connection is not a 
perfect solution, and there is still much room for further improvement.
UNet++ \cite{a:unet++} also considers reorganizing skip connections. They apply a series of 
nested dense convolutions as skip pathways to communicate between encoder and decoder sub-networks. However, they did not realize the 
importance of fully exchange information between skip connections, and in their network the information could only passed 
from deep layers up to shallow layers and the information exchange is in single direction. 
\cite{c:uctransnet} thoroughly studied the effect of each skip connection of U-Net and found that not every 
skip connection benefits the model performance. With these findings, 
they came up with UCTransNet, which is a U-Net shaped network embedded with a fusion module. The fusion module helps 
to exchange information among all feature maps produced by the encoder before conducting skip connection. In their 
CCT fusion module, the different feature maps produced by encoder layers are combined together and then channel-wise 
attention is applied to exchange information between the combined map and each single feature map.
We argue this design has two main shortcomings: 
Firstly, combining all feature maps together forms a giant feature map and it brings much greater computation cost 
to fusing operations; 
Secondly, the feature maps from very deep layers and from very shallow layers are intrinsically different, and we may not expect to 
gain much by fusing them. 

Though the designs of fusion module from previous works are defective, they inspired us for the direction 
to further improve fusion design.
And based on that we propose FusionU-Net, which contains the fusion module better handles the problems 
we've discussed above.

\section{Model Description}
\label{sec: model}
\begin{figure}
    \includegraphics[width=0.95\textwidth]{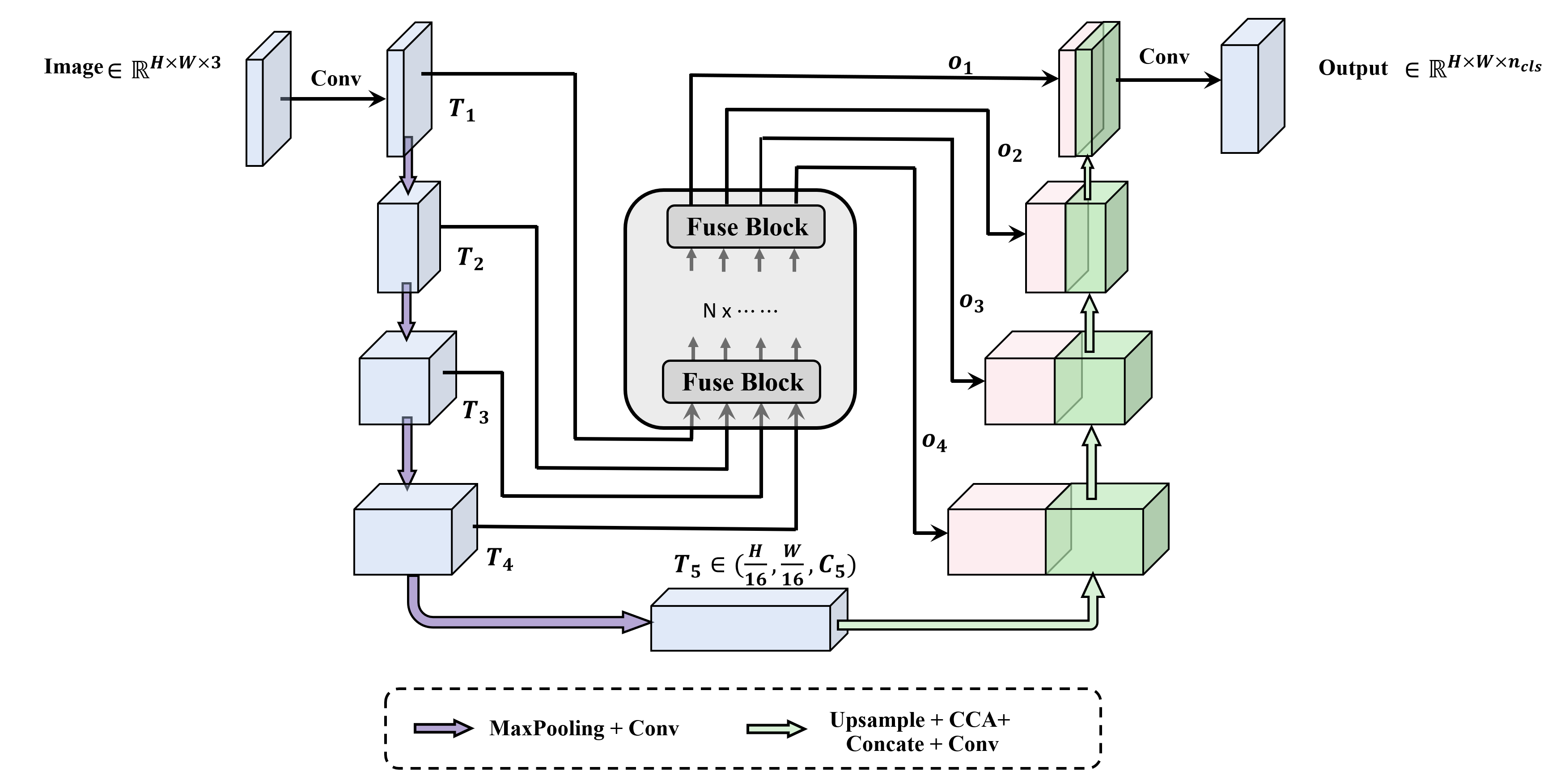}
    \caption{An overview of FusionU-Net structure}
    \label{fig:overview}
\end{figure}

In this section, we describe our proposed FusionU-Net in detail. Figure \ref{fig:overview} illustrates the overview of our model. 
In essence, our model is a U-Net-shaped network that incorporates a fusion module to strengthen the skip connections.
The whole structure can be 
divided into three parts: the encoder part, the fusion module part and the decoder part. In the following sections, 
we will thorough illustrate the details of these parts.

\begin{figure}[htbp]
    \centering
    \includegraphics[width=0.98\textwidth]{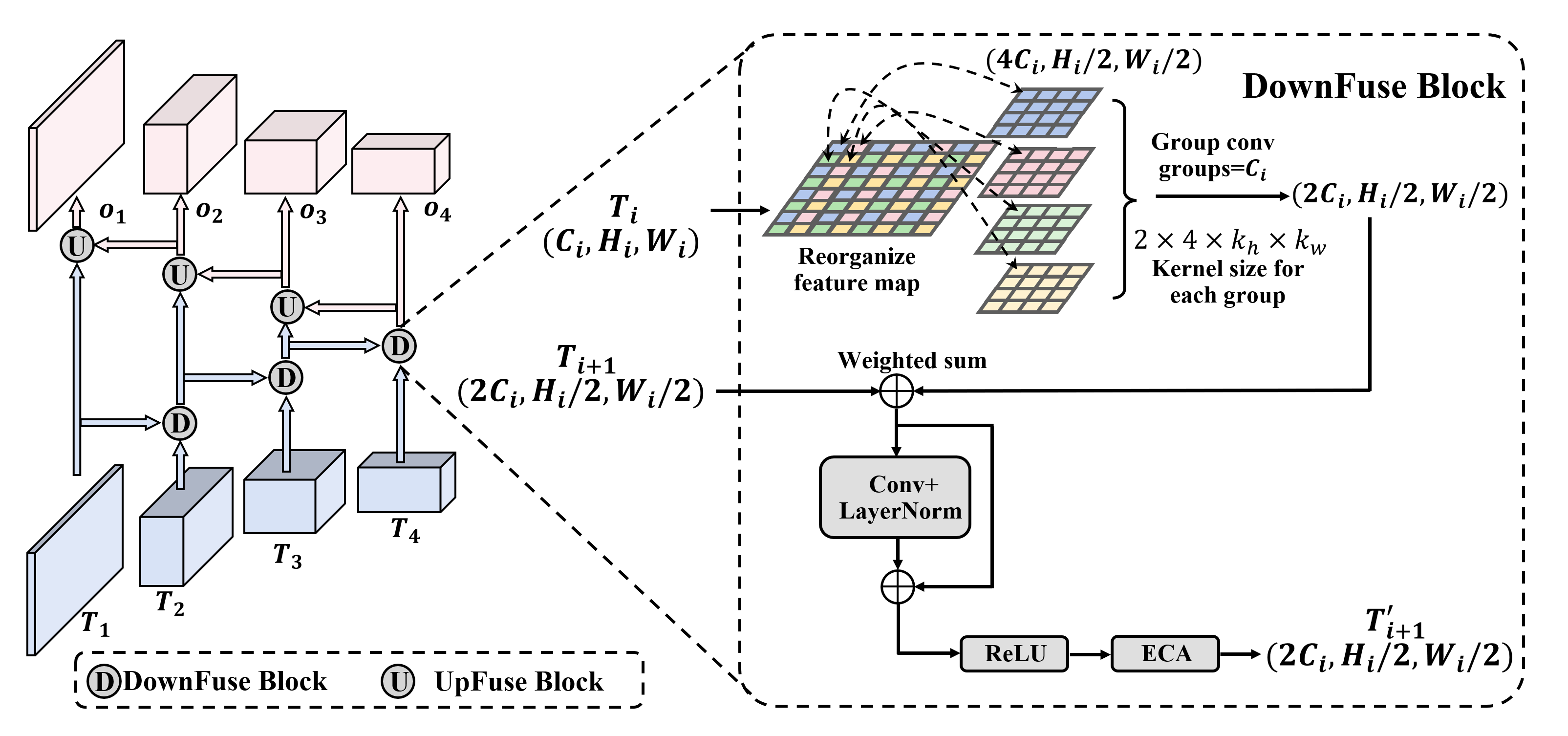}
    \caption{The structure of Fuse Block and DownFuse Block. The left part is Fuse Block structure while the 
    right part represents the details of DownFuse Block}
    \label{fig:fusion_design}
\end{figure}
\subsection{Encoder and Decoder}
To ensure fairness and facilitate future comparison with other baseline models, as well as to evaluate the effectiveness of 
our fusion module, we have adopted a similar encoder and decoder design to UNet and it is consistent with UCTransNet \cite{c:uctransnet}. 

The encoder is composed of 1 stem convolution 
block and 4 DownBlocks. Each DownBlock contains two convolution layer and a MaxPooling layer. 
For each DownBlock, the height and width of the feature map will shrink to half of its original 
size while the number of channels is doubled. The 4 outputs $T_1$, $T_2$, $T_3$, $T_4$ which are from the stem convolution block 
and the following three DownBlocks  will be passed into the Feature Fusion module for further process.

The decoder recovers the feature map from the encoder to the original size with the help of 4 UpBlocks. The UpBlock will use bilinear 
interpolation for upsampling and then apply CCA module proposed by UCTransNet to compute channel attention together with feature map 
from skip connection. After that the two feature maps will be concatenated and passed to a convolution layer. 
At the final phase, a $1\times1$ convolution is used to transfer the channel number into the number of classes 
as the final segmentation results. 

\subsection{Feature Fusion Module}
As previously discussed, the semantic information of feature maps from shallow layers and deep layers are 
vastly different. Therefore we choose to conduct feature fusion for each two feature maps from adjacent encoder outputs. 
Our fusion module is composed by stacking FuseBlocks, each FuseBlock consists of DownFuse blocks 
and UpFuse blocks as is shown in Figure \ref{fig:fusion_design}. Noticed that the structures of UpFuse block and DownFuse 
block are similar thus we only represent the DownFuse block structure here.
Both DownFuse and UpFuse blocks handle two adjacent skip connection feature maps and fuse them together. 
The entire fusion process is divided into two rounds, feature fusion is only conducted between two adjacent 
feature maps in one direction within each round. This can be either from top encoder layers to bottom encoder layers or the inverse way. 
More specifically, we firstly operate downward feature 
fusion on ($T_1$, $T_2$), ($T_2$, $T_3$) and finally ($T_3$, $T_4$), then in the opposite direction, 
we operate upward feature fusion on ($T_4$, $T_3$), ($T_3$, $T_2$) and ($T_2$, $T_1$) respectively. When 
doing feature fusion between $T_3$ and $T_4$, $T_3$ has already exchanged information with $T_2$ which has also combined 
features from $T_1$, thus information from $T_1$ could be indirectly passed to $T_4$. The second round of 
feature fusion aims to enable information to be transferred back into the shallow layers 
such that the changes of $T_4$ could be perceived by $T_1$ and this helps to achieve a bi-directional information transfer.

\subsubsection{DownFuse Block} 
The DownFuse block takes the feature map from two adjacent encoder output feature maps, 
then fuse them and produce a feature map with the same size as the one with higher channel number.
For example, we use $T_i$ and $T_{i+1}$ to represent 
the feature maps of two adjacent encoder output, and their shapes are $(C_i, H_i, W_i)$ and $(2C_i, H_i/2, W_i/2)$.
To fuse them together, we first transfer $T_i$ into
the same shape as $T_{i+1}$, then combine the two featue maps into one and apply several computing modules 
to further fuse the features. The complete working process can be described as follows:
\begin{itemize}
    \item [1] We conduct feature map reorganization to transfer $T_i$ into ($4C_i$, $H_i$/2, $W_i$/2): 
    The original feature map could be viewed as multiple 2-dimensional feature map stacked in channel-wise, 
    and for each 2-D feature map, we sample the pixels with an 
    interval of distance 1 horizontally and vertically, thus it is divided into 
    4 sub-graphs with half of original height and width. After that we stack these 4 sub-graphs at the 
    channel dimmension, then we transfer a feature map from shape ($1$, $H_i$, $W_i$) into ($4$, $H_i/2$, $W_i/2$).  
    For each channel of $T_i$, we apply above operations and then we get $T_i$ with shape 
    $(4C_i, H_i/2, W_i/2)$. The key of this design is that, unlike Pooling operations, we changed the 
    feature map size without information loss. Furthermore, we also reserve the locality of 
    previous feature map: the originally spatial neighboring 4 pixels now is still adjacent in channel-wise, 
    in other words, \emph{we transfer the spatial adjacency into channel adjacency}. 
    \item [2] We apply group convolution with number of output channels set to $2C_i$ and number of groups 
    set to $C_i$ to transfer $T_i$ into shape $(2C_i, H_i/2, W_i/2)$ which is exactly the shape of $T_{i+1}$. Notice that the kernel size of convolution
    for each group will be $2\times4\times k_h\times k_w$, thus the channel-wise pixels involved in the convolution are exactly 
    the previously four adjacent pixels, and through this way locality of feature map is still well reserved and the computation cost 
    also decreased vastly compared with normal convolution.
    \item [3] After making the two feature map to have same shape, we fuse them into one by a weighted sum. 
    \item [4] The combined feature map from stage 3 is process by a convolution layer and an ECA layer \cite{c:eca} to better 
    fuse and learn spatial and channel information.
\end{itemize}

\subsubsection{UpFuse Block}
The UpFuse block is pretty similar with the DownFuse block, we also firstly transfer the two feature maps into same shape   
and combine them together by weighted summation, then feeding the feature map into a convolution layer 
and an ECA layer. The only difference is that for UpFuse, we transfer the feature map with smaller spatial size into 
larger size by doing the inverse operation as the re-organize operation in DownFuse, and the rest procedure is 
basically the same.

\begin{table}
\resizebox{\textwidth}{!}{
\begin{tabular}{lcccccc}
    \toprule
    Models  & Params & Flops & \multicolumn{2}{c}{MoNuSeg} & \multicolumn{2}{c}{GlaS} \\ \cmidrule(lr){4-7}
    & (M) & (G) & Dice(\%) & IoU(\%) & Dice(\%) & IoU(\%) \\
    \midrule
    U-Net           & 14.75 & 25.18 & 76.51 $\pm$ 2.54 & 63.13 $\pm$ 3.15 & 86.42 $\pm$ 1.54 & 76.92 $\pm$ 1.47 \\
    U-Net++         & 9.16 & 26.72 & 77.83 $\pm$ 2.16 & 63.72 $\pm$ 2.92 & 88.02 $\pm$ 1.95 & 81.30 $\pm$ 1.50  \\
    SwinU-Net       & 27.14 & 5.91 & 77.85 $\pm$ 2.20 & 64.00 $\pm$ 1.95 & 84.84 $\pm$ 1.70 & 75.43 $\pm$ 2.15  \\
    TransU-Net      & 93.23 & 24.67 &  76.91 $\pm$ 1.95 & 62.33 $\pm$ 2.49 & 90.29 $\pm$ 0.92 & 82.97 $\pm$ 1.39  \\
    UCTransNet      & 66.24 & 32.98 & 79.17 $\pm$ 2.34 & 65.91 $\pm$ 2.85 & 89.70 $\pm$ 1.41 & 82.04 $\pm$ 2.12  \\
    \textbf{ours}   & 25.80 & 55.95 & \textbf{80.04 $\pm$ 1.38} & \textbf{66.38 $\pm$ 2.52} & \textbf{91.05 $\pm$ 1.65} & \textbf{83.23 $\pm$ 2.03} \\
    \bottomrule
    \end{tabular}}    
    \caption{The comparison with other models on MoNuSeg and GlaS datasets. For MoNuSeg and GlaS datasets we report 
    three times of five fold cross validation 
    results with form 'mean $\pm$ std'. The params and flops are calculated with input shape $1\times3\times224\times224$}
    \label {t:perform_monu_glas}
\end{table}

\section{Experiments}
\label{sec: experiments}
\subsection{Datasets}
We apply MoNuSeg \cite{a:monuseg}, GlaS \cite{a:glas} and PanNuke \cite{a:pannuke} datasets to evaluate our 
model. The MoNuSeg datasets consits of 44 images, 30 for training and 14 for testing. The GlaS dataset has 
85 images for training and 80 for testing. Compared with MoNuSeg and GlaS, PanNuke is a much more challenging task: 
PanNuke consists of exhaustible labels from 19 different tissues, and the 7904 images are randomly sampled from more 
than 20K whole slide images at different magnifications from multiple data sources.

\begin{table}
    \resizebox{\textwidth}{!}{
        \begin{tabular}{lcccccc}
            \toprule
            Models & Neoplastic & Inflammatory & Connective & Dead & Non-Neopla & Average \\
            \midrule
            UNet        & 73.09 & 56.73  & 57.12 & 19.49 & 56.78 & 52.64 \\
            UNet++      & 73.25  & 57.02 & 60.58 & 20.75 & 63.79 & 55.07 \\
            SwinUNet    & 71.57  & 54.62 & 56.41 & 19.86 & 56.48 &  51.79 \\
            TransUNet   & \textbf{76.98} & 57.77 & 61.44 & 24.24 & \textbf{68.62} & \textbf{57.81} \\
            UCTransNet  & 74.59 & 56.51 & \textbf{61.81} & 22.72 & 67.67 & 56.66 \\
            ours        & 75.37 & \textbf{58.29} & 60.71 & \textbf{24.32} & 67.78 & 57.29 \\
            \bottomrule    
        \end{tabular}
    }
    \caption{Comparison of different models on PanNuke dataset.}
    \label{t:perform_pannuke}
    \end{table}
    
    \subsection{Implementation Details}
    We train and test our model with PyTorch on a Nvidia 3090 GPU with 24 GB memory. 
    For MoNuSeg and GlaS datasets we follow the setting in \cite{c:uctransnet} and set the batch size as 4, and for 
    PanNuke dataset the batch size is set to 16. Notice that images are all resized into $224\times224$ before 
    feeding into the network. We also use simple image flip and rotation for data augmentation. As for loss functions, 
    we employ cross entropy loss and dice loss on MoNuSeg and GlaS datasets, 
    and for PanNuke, focal loss is applied to alleviate 
    the adverse impact of imbalanced distribution of different labels.
    The learning rate scheduler of cosine annealing warm-up restart is employed to avoid getting stucked in local optimal too early. 
    For MoNuSeg and GlaS, due to the limited number of training materials, we apply 
    three times 5-fold cross-validation and average 15 results of all folds as the final value to make the results 
    more convincing. For PanNuke dataset, we split the whole dataset into 
    training set, validation set and test set with ratio 7:1:2. 
    The model with best performance on validation set is chosen for final testing. 
    For all models we do not use any pretrained-weights and directly 
    initialize the weights before training the model. Dice coefficients and IoU are reported on MoNuSeg and GlaS as 
    evaluation metrics, and for PanNuke we report the mean dice and dice coefficents of all 5 labels.
    \begin{figure}
        \includegraphics[width=\textwidth]{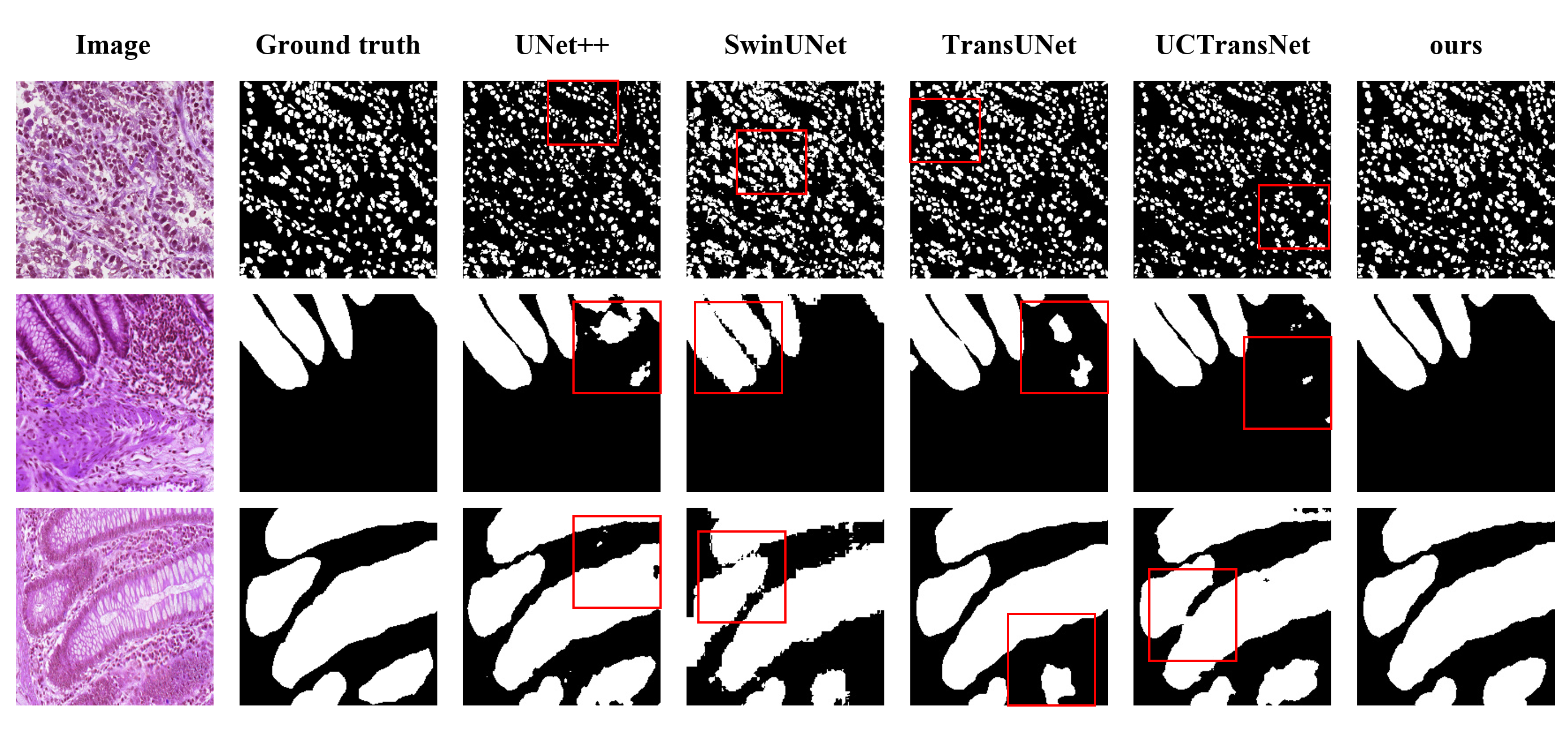}
        \caption{Segmentation results on MoNuSeg and GlaS datasets}
        \label{fig:monu_glas}
    \end{figure}

    \begin{figure}
    \begin{center}
    \includegraphics[width=1.0\textwidth]{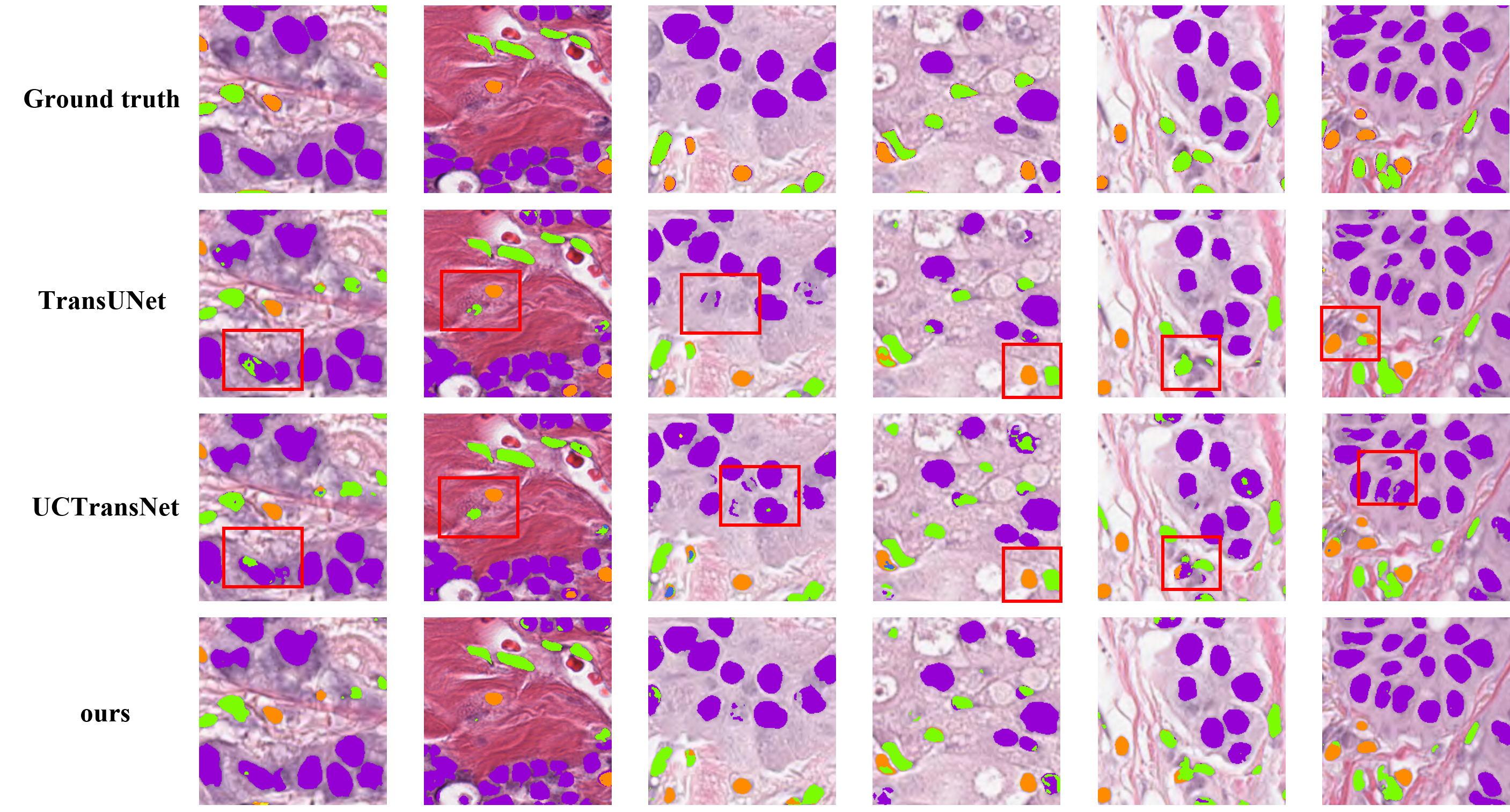}
    \caption{Segmentation results on PanNuke dataset}
    \label{fig:compare_pannuke}
    \end{center}
    \end{figure}
    
    \subsection{Comparison with other state-of-the-arts}
    To fully evaluate model performance, we compare our model with other state-of-the-art methods: UNet++ \cite{a:unet++}, 
    UCTransNet \cite{c:uctransnet}, TransUNet \cite{a:transunet} and SwinUNet \cite{c:swinunet}. 
    For fairness, the code and settings of other models are kept consistent with 
    those from publicly realeased resources. Table \ref{t:perform_monu_glas} reports the performance of all models on 
    MoNuSeg and GlaS datasets, the results strongly prove that our model outperforms the others. 
    Table \ref{t:perform_pannuke} shows the results on PanNuke datasets, from which 
    a similar conclusion could also be drawn. It is evident that our model achieves better predicting accuracies with 
    relatively less training parameters. This success can be attributed to the better handling of skip connection. 
    As discussed earlier, previous works like TransUNet and SwinUNet place too much emphasis on 
    enhancing the encoder/decoder, these designs do help to improve the accuracy but the limits of 
    naive skip connections constrain for further performance improvement. Other works like UNet++ try to enhance the skip 
    connections part, they cannot achieve fully information exchange between skip connections because of the 
    information flow is in single-direction. UCTransNet applies attention mechanism on each single and the combined feature maps to
    address that issue, but it introduces large number of extra training parameters and loses the local relevance. 
    These problems are effectively dealt with by our fusion module, as evidenced by the experiment results.
    People may notice from the table that our model has relatively higher FLOPs, but this does 
    not necessarily mean the model runs slower. We have conducted tests on the inference speed and found that our model 
    reaches 48.24 fps, which is faster than UCTransNet with 41.60 fps. The increased FLOPs are primarily attributable 
    to the dense convolution employed for fusing feature maps from deep layers, and it is our future work to  
    address this unintended consequence with improved handling.
    
    On PanNuke dataset our model also achieves favorable results, though the average dice seems to be slightly inferior to TransUNet, 
    this is mainly due to the difference of the encoder. 
    Our FusionU-Net primarily focus on fusing features across skip connections, and we utilize a relative simple encoder. 
    In contrast, TransUNet places great emphasis on the encoder design and apply dense Transformer blocks to extract information 
    on the final encoder layer, resulting in the network to be a nearly 4 times the size of our FusionU-Net. For complex tasks 
    like PanNuke, the encoder has become the bottleneck and constrained the performance of our model.
    From a different perspective, our model's ability to achieve equivalent performance with a basic encoder and limited 
    parameters proves the effectiveness of our fusion module, and this could be further supported by the comparison with other 
    models like UCTransNet and UNet++.

    \subsection{Analytical Study}
    We have also conducted ablation studies to test the effectiveness of our module design and evaluate 
    the model under various settings. These studies includes: 
    1) the contribution of DownFuse and UpFuse blocks
    2) the effectiveness of re-organize + group convolution compared with Pooling + convolution; 
    
    \begin{table}
        \centering
    \begin{tabular}{lcccccc}
        \toprule
        Method  &   \multicolumn{2}{c}{MoNuSeg} & \multicolumn{2}{c}{GlaS} & PanNuke \\ \cmidrule{2-5} 
        & Dice & IoU & Dice & IoU & (mean Dice) \\
        \midrule
        No Fusion        & 76.51 & 63.31 & 86.42 & 77.08 & 53.82 \\
        Only Downward    & 76.95 & 63.27 & 86.96 & 78.23 & 54.54 \\ 
        Only Upward Fuse & 77.87 & 64.51 & 88.42 & 79.12 & 55.75 \\ 
        DownFuse + UpFuse & \textbf{78.12} & \textbf{64.73} & \textbf{89.73} & \textbf{80.53} & \textbf{57.29} \\
        \bottomrule
    \end{tabular} 
    \caption{Ablation study about the effectiveness of DownFuse block and UpFuse block}
    \label{t:ab_fuseblock}
    \end{table}
    \begin{table}
        \centering
    \begin{tabular}{lcccccc}
        \toprule
        Method  &   \multicolumn{2}{c}{MoNuSeg} & \multicolumn{2}{c}{GlaS} & PanNuke \\ \cmidrule{2-5} 
        & Dice & IoU & Dice & IoU & (mean Dice) \\
        \midrule
        Pooling+Conv                 & 77.49 & 63.81 & 88.95 & 80.02 & 56.89 \\
        Reorganize+Group-Conv        & \textbf{78.12} & \textbf{64.73} & \textbf{89.73} & \textbf{80.53} & \textbf{57.29} \\
        \bottomrule
    \end{tabular} 
    \caption{Ablation study about the effectiveness of reorganize + group convolution}
    \label{t:ab_reorganize}
    \end{table}

\paragraph{Contribution of DownFuse and UpFuse:}
To prove the value of two-round fusion, we change the originial FuseBlock into one round fusion with either 
DownFuse or UpFuse and we also test for directly using the raw skip connection to show the 
necessity of doing feature fusion(notice we still use CCA for decoder part feature fusion under these settings). 
The experiment result is presented in table \ref{t:ab_fuseblock}, and it demonstrates that the fusing operation 
before the skip connections 
is a useful way to enhance model performance. And we could observe that applying upward 
fusion helps more than downward fuse. This is because the original U-Net encoder forward process can also 
be viewed as a similar progress with our downward fusion, thus the upward feature fusion which assists in 
passing the message from bottom back to top layers enable the model to better exchange information between layers. 
And it is also worth pointing out that even though downward fuse helps little alone, combining it with 
upward fuse will benefit a lot to the model just as the data shows.

\paragraph{Effectiveness of Reorganize and Group-conv:}
To keep the local adjacency and avoid information loss during the downsampling and upsampling process, we propose 
a better alternative by reorganizing pixels of the feature map and applying group convolution. To show the advantage of 
this method, we substitute the reorganize and group convolution with Pooling and a normal convolution and test 
the performance on three datasets. The new 
model's number of training parameters and flops grows into 34.98M and 97.15G, while the segmentation performance 
dropped as in shown in table \ref{t:ab_reorganize}.  
Therefore, it is evident that our design is superior to the traditional way.

\section{Conclusion}
U-Net is a highly effective model for pathology image segmentation and has inspired numerous works. 
In this paper, we aim to explore a different perspective on improving segmentation accuracy by modifying the 
original skip connection design. Our proposed network, FusionU-Net, is similar to U-Net but with an additional 
fusion module that applies feature fusion on skip connections. The fusion module is based on a two-round design 
to better handles the relevance of adjacent encoder outputs and enables information to be fully exchanged 
between any two skip connection feature maps. Additionally, we propose a new way of upsampling and downsampling 
by reorganizing the feature map and applying a group convolution. This approach can avoid information loss and 
preserve the pixel adjacency relationship effectively, which is particularly suitable for 
pathology images. Through extensive experiments and in-depth analysis, we demonstrate that our model is highly 
effective and our fusion module represents a superior method for fusing feature maps from skip connections.

\section*{Acknowledgments}
This work was supported in part by the National Natural Science 
Foundation of China under Grant 62101318, and the Key Research and 
Development Program of Jiangsu Province, China under Grant BE2020762.

\bibliographystyle{unsrt}  
\bibliography{references}

\end{document}